\documentclass[twocolumn,showpacs,preprintnumbers,amsmath,amssymb,superscriptaddress]{revtex4}
\usepackage{graphicx}
\usepackage{amsmath}
\usepackage{amssymb}
\usepackage{here}

\begin{document}
\title{Phase competitions in epitaxial 
Pr$_{0.5}$Ca$_{0.5}$MnO$_3$/La$_{0.5}$Sr$_{0.5}$MnO$_3$ 
superlattices}
\author{H.~Wadati}
\email{wadati@ap.t.u-tokyo.ac.jp}
\homepage{http://www.geocities.jp/qxbqd097/index2.htm}
\affiliation{Department of Applied Physics and Quantum-Phase Electronics 
Center (QPEC), University of Tokyo, Hongo, Tokyo 113-8656, Japan} 
\author{J. Okamoto}
\affiliation{Condensed Matter Research Center and Photon Factory,
Institute of Materials Structure Science, High Energy Accelerator
Research Organization, Tsukuba 305-0801, Japan}
\author{M. Garganourakis}
\affiliation{Swiss Light Source, Paul Scherrer Institut, 
5232 Villigen PSI, Switzerland} 
\author{V. Scagnoli}
\affiliation{Swiss Light Source, Paul Scherrer Institut, 
5232 Villigen PSI, Switzerland} 
\author{U. Staub}
\affiliation{Swiss Light Source, Paul Scherrer Institut, 
5232 Villigen PSI, Switzerland} 
\author{M. Nakamura}
\affiliation{RIKEN Center for Emergent Matter Science (CEMS), Wako 351-0198, Japan}
\author{M. Kawasaki}
\affiliation{Department of Applied Physics and Quantum-Phase Electronics 
Center (QPEC), University of Tokyo, Hongo, Tokyo 113-8656, Japan} 
\affiliation{RIKEN Center for Emergent Matter Science (CEMS), Wako 351-0198, Japan}
%\affiliation{Cross-Correlated Materials Research Group (CMRG), 
%RIKEN Advanced Science Institute, Wako 351-0198, Japan}
\author{Y. Tokura}
\affiliation{Department of Applied Physics and Quantum-Phase Electronics 
Center (QPEC), University of Tokyo, Hongo, Tokyo 113-8656, Japan} 
\affiliation{RIKEN Center for Emergent Matter Science (CEMS), Wako 351-0198, Japan}
%\affiliation{Cross-Correlated Materials Research Group (CMRG), 
%RIKEN Advanced Science Institute, Wako 351-0198, Japan}
\date{\today}
\begin{abstract}
We studied the charge-orbital ordering 
in the superlattice of charge-ordered 
insulating Pr$_{0.5}$Ca$_{0.5}$MnO$_3$ 
and ferromagnetic metallic La$_{0.5}$Sr$_{0.5}$MnO$_3$ 
by resonant soft x-ray diffraction. 
A temperature-dependent incommensurability is found 
in the orbital order. 
In addition, a large hysteresis is observed that is caused 
by phase competition between insulating charge ordered 
and metallic ferromagnetic states. No magnetic phase transitions are 
observed in contrast to bulk, confirming the unique 
character of the superlattice. 
The deviation from the commensurate orbital order 
can be directly related to the decrease of ordered-layer thickness 
that leads to a decoupling of the orbital-ordered planes along the $c$ axis. 
\end{abstract}
\maketitle
%\section{Introduction}
Hole-doped perovskite manganites $RE_{1-x}A_x$MnO$_3$, 
where $RE$ is a rare-earth ($RE=$ La, Nd, Pr) and $A$ is 
an alkaline-earth atom ($A=$ Sr, Ba, Ca) have 
attracted much attention because they exhibit 
remarkable physical properties such as colossal 
magnetoresistance (CMR) and complex 
electronic ordering phenomena 
\cite{rev,RamirezMn,orbital,RaoMn,PrellierMn,
DagottoMn,Hungry,TokuraMn}. 
La$_{1-x}$Sr$_{x}$MnO$_3$ has a 
large bandwidth and a ferromagnetic metallic (FM) 
ground state is realized for approximately 
$0.2<x<0.5$ \cite{urushibara}. 
%whereas for $x>0.5$, this phase competes 
%with a charge and orbitally ordered (CO/OO) 
%insulating antiferromagnetic (AF) phase \cite{urushibara}. 
Most of the half-doped manganites ($x\simeq 0.5$) 
with a small bandwidth exhibit the so-called ``CE-type'' 
charge and orbitally ordered (CO/OO) insulating and 
antiferromagnetic (AF) ordering 
with alternating Mn$^{3+}$ and Mn$^{4+}$ states 
within the (001) plane in the form 
of stripes \cite{JirakPCMO}. 
This ordering competes with the FM phase. 
Phase competition between ordered phases is 
an interesting phenomenon 
and leads to intriguing behaviors such as CMR 
and nanometer-scale phase separation. 
These ordering phenomena lead to symmetry 
lowering and a doubling of the unit cell, 
which results in superlattice reflections 
observable with different scattering techniques. 
For higher doping levels $x>0.5$, the doping 
leads to orderings, that are incommensurate, 
with ordering wave vector proportional 
to the doping concentration 
\cite{little,Shimomura}. 

In recent years, epitaxially grown films 
of manganites have been extensively studied. 
It was found that La$_{1-x}$Sr$_x$MnO$_3$ 
thin films are ferromagnetic 
for approximately $0.2<x<0.5$, almost 
the same as the bulk \cite{Izumi,horibaLSMO}. 
Half-doped manganite thin films 
remain often charge and orbitally ordered, 
but physical properties might depend on the strain, that is, 
the lattice constant and orientation of the substrates. 
A transition between CO and FM states was observed 
in Nd$_{0.5}$Sr$_{0.5}$MnO$_3$ thin films 
on SrTiO$_3$ substrates only when the 
substrate orientation was (011) \cite{NSMONakamura,Waka}. 
Also in Pr$_{0.5}$Ca$_{0.5}$MnO$_3$ thin films, 
epitaxial strain strongly affects the electronic properties, 
and the thin films grown epitaxially 
on (LaAlO$_3$)$_{0.3}$(SrAl$_{0.5}$Ta$_{0.5}$O$_3$)$_{0.7}$ (LSAT) 
(011) substrates exhibit a CO transition around 220 K, 
similar to bulk samples \cite{okumura}.

Interesting phenomena occur when these two systems are brought 
in direct contact. Results on such stackings has recently been reported by
Nakamura {\it et al.} \cite{SL}. 
They fabricated superlattices of 
FM La$_{0.5}$Sr$_{0.5}$MnO$_3$ (LSMO) and 
CO Pr$_{0.5}$Ca$_{0.5}$MnO$_3$ (PCMO) on LSAT 
(011) substrates \cite{SL}. 
They found that they could control the phase boundary 
at the interface between the ferromagnetic LSMO and 
the AF and CO/OO PCMO 
by the application of magnetic fields or 
changing the individual layer thicknesses. In addition, 
they observed superlattice reflections indicative 
for the structural distortions induced by the charge and orbital 
order in these systems. Particularly interesting is 
the case of having equal thicknesses. In that case 
the reflections exhibit a strong hysteresis behavior 
even in zero magnetic field, 
a direct evidence for the first-order phase 
transition between these competing phases. 

To gain more insight to these problems 
of phase transitions and phase competitions, 
we investigated the charge-orbital ordering 
in [PCMO (5 layers)/LSMO (5 layers)]$_{15}$ 
superlattices by resonant soft x-ray diffraction. 
This technique combines the sensitivity of the Mn $2p\rightarrow 3d$ 
electronic transition to the electronic and magnetic properties 
of the transition metal $3d$ states with the sensitivity 
to long range order of diffraction. 
It has been used to disentangle magnetic and orbital ordering phenomena 
in manganites \cite{LSMORXS2,LSMORXS,dhesi,kjt,Wilkinsnjp,
StaubPCMONSMO,Garcia2,ALSPCMO,WadatiYMO,MariosPRL,HWcond}, 
and is extremely sensitive to even small details 
in the electronic orderings. 
It is especially sensitive to the magnetic structures in thin films 
\cite{valerio,WadatiYMO,MariosPRL,HWcond}, 
and has recently been used to study effect of imprinting magnetic and 
electronic information in epitaxially grown PCMO films \cite{MariosPRL}. 
In the present study we find a significant hysteresis behavior of 
the superlattice reflection that is sensitive to the orbital order 
of the Mn $3d$ states. In addition the system shows a clear 
temperature dependence of the ordering wave vector 
with hysteresis behaviors. 
The temperature dependent incommensurability in the orbital ordering 
can be related to the change of orbital coupling 
through the FM metallic layer centered around the LSMO layer. 

\begin{figure}
\begin{center}
\includegraphics[width=4cm]{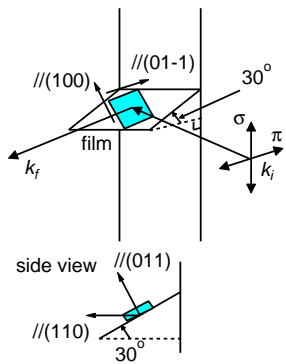}
\caption{(Color online): 
The experimental geometry having the [110] 
axis of the PCMO/LSMO superlattice 
in the scattering plane.}
\label{geo}
\end{center}
\end{figure}

%\section{Experiment}
[PCMO (5 layers)/LSMO (5 layers)]$_{15}$ superlattices were 
grown on the $(011)$ surface of an LSAT substrate 
by pulsed laser deposition. 
The details of the sample fabrication 
were described elsewhere \cite{SL}.
Resonant soft x-ray diffraction experiments were performed
on the RESOXS end station \cite{SLS} at the surface-interface
microscopy (SIM) beam line \cite{aip} of the Swiss
Light Source of the Paul Scherrer Institut, Switzerland. 
The experimental geometry is shown in Fig.~\ref{geo}, which is the same 
as that in Ref.~\cite{HWcond}. A continuous helium-flow cryostat 
allows measurements between 10 and 300 K. 
The experiments were performed in the out-of-focus mode 
with a beam of approximately $2\times 2$ mm$^2$. 

\begin{figure}
\begin{center}
\includegraphics[width=9cm]{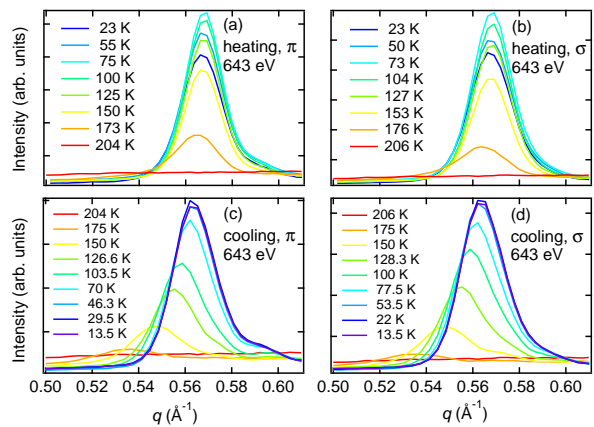}
\caption{(Color online): 
Temperature dependence of the 
$(h h 0)$ peak. 
Panels (a) and (b) ((c) and (d)) were 
measured in the heating (cooling) cycle. 
Incident x-ray polarizations were 
$\pi$ in (a) and (c) and 
$\sigma$ (b) and (d). 
All the data were taken at $h\nu=643$ eV 
(Mn $2p_{3/2} \rightarrow 3d$ 
absorption edge). }
\label{fig1}
\end{center}
\end{figure}

\begin{figure}
\begin{center}
\includegraphics[width=9cm]{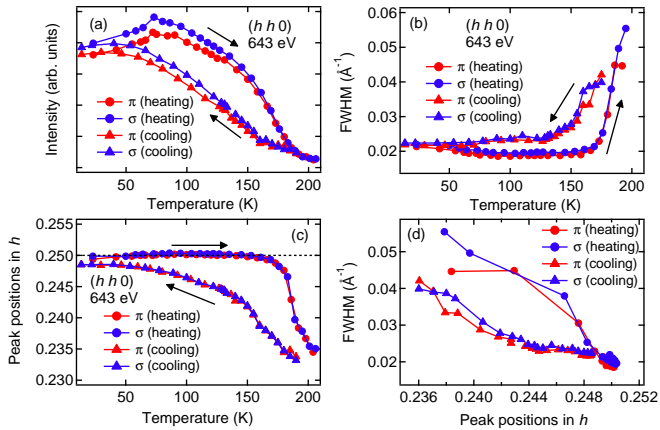}
\caption{(Color online): 
Temperature dependence of the 
$(h h 0)$ peak 
intensity (a), 
width (b), and 
positions (c). 
In panel (d), the peak width is 
plotted as a function of 
peak positions. }
\label{fig2}
\end{center}
\end{figure}

%\section{Results and Discussions}
Figure \ref{fig1} shows the $q$ dependence 
of the $(h h 0)$ reflection 
for various temperatures at the Mn $2p_{3/2}$ edge 
(643 eV) for both $\pi$ and $\sigma$ 
incoming x-ray polarizations. This reflection has been studied 
in detail in Pr$_{1-x}$Ca$_x$MnO$_3$ bulk manganites and epitaxial grown films 
\cite{kjt,StaubPCMONSMO,ALSPCMO,MariosPRL,HWcond}, 
and has been shown to be sensitive to 
both the orbital and magnetic orderings of the systems. 
The temperature dependence of this reflection 
in a heating run (panels (a) and (b)) significantly deviates 
from that in a cooling run (panels (c) and (d)). 
The reflection appears below $\sim 200$ K, 
which is slightly lower than the CO 
transition temperature of 220 K in the pure 
PCMO thin film \cite{okumura}. 
There is almost no difference 
between $\pi$ (a, c) and 
$\sigma$ (b, d) polarizations. 
The temperature variation 
of the corresponding peak intensity, 
the full width at half maximum (FWHM), 
and positions is summarized 
in Fig.~\ref{fig2}. 
In all of these quantities, 
a large hysteresis is observed between cooling and heating cycles. 
The peak intensity monotonically increases for decreasing temperatures 
in contrast to the heating cycle, where, interestingly, 
it first increases and then decreases before it disappears 
around 200 K. This is related to both the temperature dependence 
of the position and FWHM of the reflection. 
The peak position is incommensurate and 
temperature dependent in the full range of 
the cooling cycle. In contrast, 
the position remains frozen until temperatures of 
approximately 170 K for the heating cycle. 
As the peak position of this reflection type has been found 
to be related with the doping level \cite{Garcia2}, 
it might reflect an {\it effective} doping in the PCMO layer, 
which is slightly larger than 0.5 and is temperature dependent. 

%This is contrasted by the cooling cycle, where the position 
%is monotonically increasing for decreasing temperatures. 
The width also shows a an interesting behavior. 
When the FWHM is minimal, 
the intensity is maximal. 
In the heating cycle, this is the case 
for temperatures between 50 and 150 K. 
This shows that there is a clear connection 
among the improved order (maximal intensity), 
the correlation (maximal correlation length), 
and the largest $h$. 
To test if this relation is quantitatively 
the same for cooling and heating cycles, 
we show in panel (d) the FWHM 
as a function of peak position. 
Although the width is linearly dependent 
on the peak position 
for both the cooling and the heating cycles, 
the slope of the two is significantly different. 
Since the hysteresis behavior in all three parameters 
between heating and cooling cycles 
must be caused by dynamics in orbital and magnetic order, 
the same origin is expected to be responsible for 
differences in the slope. 

\begin{figure}
\begin{center}
\includegraphics[width=9cm]{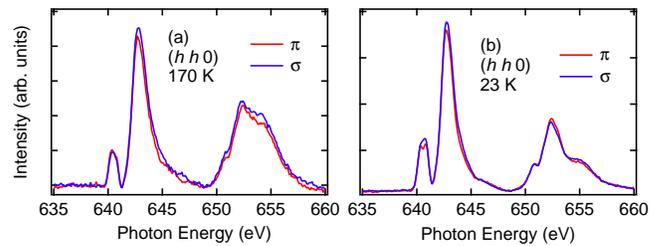}
\caption{(Color online): 
Intensity of the $(h h 0)$ peak 
as a function of photon energies 
at the Mn $2p\rightarrow 3d$ 
absorption edge at 170 K (a) and 23 K (b).}
\label{fig3}
\end{center}
\end{figure}

Figure \ref{fig3} shows the spectral shape of 
the $(h h 0)$ reflection 
in the vicinity of the Mn $2p$ 
edges at 170 K (a) and 23 K (b). 
There is no polarization dependence at both temperatures. 
In addition, the spectral shape is 
identical at these two temperatures and 
very similar to the pure orbital scattering 
of the analogue reflection 
in bulk La$_{1.5}$Sr$_{0.5}$MnO$_4$ 
\cite{LSMORXS,dhesi,Wilkinsnjp} 
and the pure PCMO thin film \cite{HWcond}. 
In other words, we find 
no indication of magnetic scattering 
in the superlattice, 
in contrast to the pure PCMO thin film \cite{HWcond}. 
There, the resonant soft x-ray diffraction revealed 
two magnetic transitions; $T_N=150$ K and $T_2=75$ K. 
The magnetic intensity of the $(h h 0)$ reflection 
in the film and the bulk 
can be directly related 
to a spin canting of the Mn spins along the $c$ axis, 
which is clearly absent in the superlattice. 
Nevertheless, the intensity maximum in the heating cycle 
occurring around 70 K coincides with the magnetic transition 
of the pure PCMO thin film. In bulk (La,Pr,Ca)MnO$_3$ systems, this 70 K 
magnetic transition is much more pronounced \cite{Wunmat} 
and is believed to be caused by a phase separation 
between FM and AF phases 
with glassy character at low temperatures. 
Such a phase transition would also be 
directly visible 
in the observed orbital reflection intensities. 
The difference between the PCMO 
thin film and the corresponding bulk material 
indicates that the epitaxial strain affects 
the magnetism in thin films \cite{HWcond}. 
The observation of the 70 K transition 
in the orbital signal in the superlattice 
is an indirect evidence that the superlattice is AF 
with spins lying fully in the $ab$-plane, 
as observed by neutron diffraction 
on PCMO bulk samples \cite{JirakPCMO}. 

\begin{figure}
\begin{center}
\includegraphics[width=9cm]{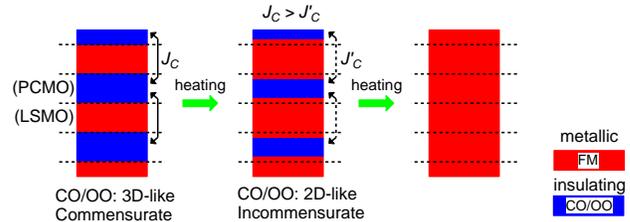}
\caption{(Color online): 
Real-space sketches of phase separations 
between FM and CO states. }
\label{fig4}
\end{center}
\end{figure}

Combining our data with the results form Ref.~\cite{SL} 
leads now to the following picture. 
The proposed phase separation 
between the FM LSMO layers and the orbital ordered 
AF PCMO layers is supported by the observation of both 
the ferromagnetic moment \cite{SL} and 
the orbital reflection. 
The sharp reflection with a correlation length 
of the order of the total film thickness 
indicates that the correlation 
between the CO/OO states perfectly 
bridges the FM layers built in between. 
This is visualized in Fig.~\ref{fig4}. 
This orbital coupling bridging over the 
FM LSMO layers becomes weak when 
the temperature increases because 
the interface CO/OO states 
disappear first, leading to 
an effective increase of 
the separation of the CO/OO 
layers (see Fig.~\ref{fig4}). 
The proximity effect leads 
to a slight change of the overall 
doping in the CO/OO layers, which 
can explain the change in the ordering 
wave vectors. 
This affects also the correlation length. 
This behavior is similar to that 
found for $2/3$ doped 
Tb$_{0.66}$Ca$_{0.33}$BaMn$_2$O$_6$ 
\cite{Garcia2}, where, due to the 
A-site order in layers, a similar 
two-dimensional (2D) character 
exists in the bulk 
(though with much shorter distances). 
Also in these systems, 
a competition between FM/AF ordering 
exists, and for a wide range of doping 
a change of ordering vector and correlation 
length has been interpreted in terms of 
a decoupling between the orbital-ordered layers 
along the $c$-axis. 

%\section{Summary}
In summary, we performed resonant soft x-ray diffraction 
measurements on the superlattice of 
charge-ordered insulating PCMO
and ferromagnetic metallic LSMO 
to study the charge-orbital ordering in 
PCMO layers. 
We found clear differences 
from the pure PCMO and LSMO thin films. 
A large hysteresis was observed 
due to the phase competition and 
different magnetic ordering 
in the superlattice. 
We found distinct changes of the 
ordering wave vector and the 
correlation length connected to 
2D characters of the system. 
These observations give 
indication for an ``effective'' 
doping caused by 
the proximity effect, 
resulting in a model of 
phase separations between 
FM and CO/OO states. 

%\section*{Acknowledgments}
The authors thank the experimental support
of the X11MA beam line staff. Financial support of the
Swiss National Science Foundation and its NCCR MaNEP
is gratefully acknowledged. This research is also granted
by the Japan Society for the Promotion of Science (JSPS)
through the ``Funding Program for World-Leading
Innovative R\&D on Science and Technology (FIRST
Program),'' initiated by the Council for Science and
Technology Policy (CSTP). 

\bibliography{LVO1tex}

\end{document}